**AI Human Impact**
*Toward a Model for Ethical Investing in AI-Intensive Companies*

## Abstract

Does AI conform to humans, or will we conform to AI? An ethical evaluation of AI-intensive companies will allow investors to knowledgeably participate in the decision. The evaluation is built from nine performance indicators that can be analyzed and scored to reflect a technology's human-centering. The result is objective investment guidance, as well as investors empowered to act in accordance with their own values. Incorporating ethics into financial decisions is a strategy that will be recognized by participants in environmental, social, and governance investing, however, this paper argues that conventional ESG frameworks are inadequate to companies that function with AI at their core. Fully accounting for contemporary big data, predictive analytics, and machine learning requires specialized metrics customized from established AI ethics principles. With these metrics established, the larger goal is a model for humanist investing in AI-intensive companies that is intellectually robust, manageable for analysts, useful for portfolio managers, and credible for investors.

## Keywords

AI, Finance, Ethical investing, AI ethics, ESG

James Brusseau
Philosophy
Pace University, New York City
jbrusseau@pace.edu
orcid.org/0000-0002-9393-7788
linkedin.com/in/james-brusseau

**Conflict of interest**: The author declares that he has no conflict of interest.
**Ethical approval**: The author declares that the paper is compliant with ethical standards.

## ESG Does Not Work for AI

Environmental, Social, and Governance (ESG) investment rating was forged for the industrial economy with its hazardous working conditions and polluting machines (MSCI 2020). Artificial Intelligence companies do not fit in. Part of the disconnect is material – cement and smokestacks diverge from pixels and digital exhaust – but the significant difference is human. When Henry Ford promised customers they could have any color they wished so long as it was black, he was not proposing a color acceptable to every purchaser so much as eliminating individuality from purchasing. Ford did *not want to know* about customers' unique preferences. The personal information was even counterproductive because making vehicles profitably depended on construing humans as monochromic and interchangeable, like the units rolling off the assembly line.

The big data and the predictive analytics economy reverses the logic, it *runs* on personalization. Netflix does not aspire to generic movie recommendations for homogenized demographic groups, it aims specific possibilities toward individual viewers at targeted moments (Guadiana 2020). The burgeoning field of dynamic insurance does not cover population segments over extended durations, it customizes for unique clients and intervenes at critical junctures (Keller 2018). AI coronary healthcare is less concerned with a patient's age cohort than with tiny and personal heartbeat abnormalities that escape human eyes but not machine-learned analysis (Zicari *et al*. 2021). In every case, identifying personal information shifts to the center of innovation.

The shift explains why privacy concerns have surged in public conversations and corporate meeting rooms (West 2019). It also means that within the scope of AI and human interaction, the most tangible peril is not captured well by standard ESG criteria (Mascotto 2020). It is not measured as environmental toxins or perilous workplaces or institutional corruption or poverty, instead, the risk is *our own* dataset, it is the information defining who we are – our habits, anxieties, beliefs, and desires – that may be engineered to provide gratifying experiences and opportunities, but that can also be twisted to control where we go and what we do.

The paradigmatic case is predictive policing (Kaufmann *et al.* 2019) because of the question it asks: *Is my data liberating, or confining?* Will the personal information that has been gathered about me invigorate my life, or restrict it? Whether the AI is stationed at a security kiosk, on the LinkedIn career platform, or the Tinder romance site, or inside a hospital emergency room, or underlying a mortgage loan decision, or behind the screen of an Amazon purchase recommendation, the question is the same.

Because the question about whether AI serves humanity, or humanity serves AI fundamentally asks whether data and algorithms vitalize or debilitate on the level of individuals, the first metric for responsible investing is autonomy: Does a technology expand self-determination? The personal values of dignity and privacy naturally cohere as key performance indicators, and it follows that what makes AI humanism different from ESG investing – and what requires a new and distinct model for ethical participation in AI-intensive companies – is evaluation that begins with unique persons, not demographic segments or collectives and their interests.

Finally, *AI Human Impact* and *Environmental Social Governance* are both nonfinancial investment strategies, but they function along different economic lines. ESG criteria suit traditional companies in the broader industrial economy, while AI Human Impact has the narrower purpose of navigating the unprecedented economic powers, social effects, and ethical dilemmas rising alongside companies that function with AI at their core.

## Overview of AI Human Impact

Much of what will be required to produce a human-centered investing blueprint for AI-intensive companies has already been accomplished. Since 2017, more than 80 AI and big data ethical principles and values guidelines have been published (Jobin *et al.* 2019: 3; Fjeld *et al.* 2020; Hagendorff 2020), and even as the wave crested, researchers were already mapping the overlaps, forming principles of the principles (Floridi and Cowls 2019, Hagendorff 2020). Because many of the proposals are ethical in origin, the contributions tend to break along the lines of academic applied ethics (Mittelstadt 2019). Corresponding with libertarianism there are values of personal freedom. Extending from a utilitarian approach there are values of social wellbeing. And, on the subject of technical accountability, there are values focusing on trust for what an AI does.

Each of the mainstream collections of AI ethics has their own way of fitting onto that trilogical foundation, but the *Ethics Guidelines for Trustworthy AI* sponsored by the European Commission (AIHLEG 2019) is representative (Clement-Jones 2019). So too is the *Opinion of the Data Ethics Commission* of Germany (DEC 2019). They are arranged in the figure below for comparison, along with the values grounding AI Human Impact: Autonomy, Dignity, Privacy in the personal freedom group; Fairness, Solidarity, Sustainability in the social wellbeing group; Performance, Safety, Accountability in the technical trustworthiness group.

*Figure 1*

| | AI Human Impact | ***Ethics Guidelines for Trustworthy AI,* European Commission** | ***Opinion,* German Data Ethics Commission** |
|---|---|---|---|
| **Personal Freedom** | Autonomy | Human agency and oversight | Self-determination |
| | Dignity | | Human dignity |
| | Privacy | Privacy and data governance | Privacy |
| **Social Wellbeing** | Fairness | Diversity, non-discrimination, fairness | |
| | Solidarity | | Justice and Solidarity |
| | Sustainability | Societal and environmental wellbeing | Sustainability |
| | | | Democracy |
| **Technical Trustworthiness** | Performance | | |
| | Safety | Technical robustness and safety | Security |
| | Accountability | Accountability | |
| | | Transparency | |

There are discrepancies, and some are superficial. The E.C. *Guidelines* split 'Accountability' and 'Transparency,' whereas AI Human Impact unites them into a single category. The German Data Ethics Commission joins 'Justice' and 'Solidarity,' whereas AI Human Impact splits them into 'Fairness' and 'Solidarity.' Another difference is more profound. Performance as an ethical principle occurs only in the AI Human Impact model because it is extremely important to investors: a primary reason for involvement in the evaluation of AI-intensive companies is the financial return. How well the technology performs, consequently, is critical because an AI that cannot win market share will not have financial, human, or any impact.

There is also a second effect stemming from this project's grounding in the finance industry. It is that ethical investing will not constrain computer scientists, engineers, or technological progress, instead, it provides humanistic feedback to *catalyze more and faster innovation*. This is not a defense *against* AI but a contribution *to* it. In at least three ways, ethics supports technological growth. First, AI Human Impact illuminates scenes of potential social objection, including unfair outcomes resulting from biased training data as may occur when employment, lending or healthcare algorithms distribute opportunities unequally (Bogen 2019). Foreseeing reputational and legal risks contributes to their avoidance. Second, irresolvable dilemmas are mapped so that they can be capably navigated, such as the facial recognition tradeoff between the right to privacy and the urge for personalized services (Press release 2019). While there may be no escape from this tension, there is a difference between controlling the dilemma and being controlled by it. The critical element of control is understanding what is at stake when values tradeoffs get made. Ethics provides the understanding. Finally, and most significantly, a humanist evaluation orients AI design toward individual potential by describing how data and machine learning can be *measurably* converted into vital experiences that replace the numbing and banal activities now consuming too many human hours. Elevating, accelerating, and multiplying personal opportunities, that is the ethical purpose of AI impact investing. The accompanying financial premise is that the humanist purposiveness yields outstanding returns.

The following sections develop nine categories for analysis and scoring. Each one reveals a discrete aspect of ethical exposure and opportunity for AI-intensive companies.

## Personal Freedom Metrics: Autonomy, Dignity, Privacy

**Autonomy** means giving rules to oneself, which conceptually stands between living under rules imposed by others, and the senseless chaos of life without any rules at all. In lived AI experience, autonomy exists when data and algorithms help us act for our own reasons.

Because no company currently reports on their autonomy bottom line, this performance indicator proves difficult to abstract from corporate disclosures and public information sources. Still, a resourceful analysis can distinguish those AI companies that constrict users' self-determination, from those that expand it.

Constrictors may employ *dark patterns*, interfaces surreptitiously prompting decisions that users might not otherwise make (Narayanan et al 2020). As Facebook's founding President explained:

> How do we consume as much of your time and conscious attention as possible? We need to give you a little dopamine because someone Liked or commented on a photo or a post, and that's going to get you to contribute more content, and that's going to get you more Likes and comments. It's a feedback loop exploiting a vulnerability in human psychology (Allen 2017).

The psychological vulnerability is chemical, and activated by social media *Likes* dosed by algorithms. The result is insidious. Users are confined by *their own data*, and confined in two senses. First, the material that keeps drawing them back is their own posting. Second, the dopamine is calibrated to the users' personal profile: if the Likes are too many for that specific person, or too few, interest dissipates (Remia 2015). In the end, it may be that users *enjoy* being constrained by their own posts and personal information, but the autonomy score is negative.

Another dark pattern is AI nudging, defined as controlling behavior without relying on legal or regulatory mechanisms (Thaler and Sunstein 2008). This behavioral modification has been modelled for use in home chatbots to "foster empathy that nudges a user towards performing charitable acts" (Borenstein and Arkin 2017: 502), with the key being information gathering and processing for surgically precise appeals. "If the user has a family history of a particular illness like heart disease, the robot could suggest contributing to an associated charity, like the American Heart Association" (Borenstein and Arkin 2017: 503). So, a sad episode converts into a beneficial donation. But the public benefit cannot be allowed to obscure what is happening in the silo of autonomy: users are not receiving data to help them make decisions so much as receiving decisions generated by their data. The AI may be presented as altruism, but its ethics score for self-determination is negative.

Autonomy can also be constricted through dependence (AIHLEG 2019: 16). In medical domains including ophthalmology, oncology and dermatology, machine learning algorithms have outperformed human counterparts in competitive tests of detecting diseases from clinical images and, in these contexts, it may become difficult for doctors to trust their own learning more than the machine's (Grote and Berens 2019). Subsequently, deference to the statistically proven performance may become habitual: if the AI is going to be right, why bother to think through a diagnosis independently? The result is deskilling, dependence, and an adverse effect on self-determination.

How does an autonomy score turn positive? With AI designed to facilitate human *doing*. While it is true that Facebook can trap users in their own banality, the platform's multiple applications and vast knowledge of its own users empowers entrepreneurs to establish a business and advertise online to well-curated potential clients in hours. A business can literally become successful overnight. And, while AI image analysis may render doctors redundant, it can also heighten performance by drawing attention to anomalies they may have otherwise missed, or by freeing time to pursue higher-level research. With respect to any specific AI-intensive company, the root autonomy question – *Does the AI open opportunities or close them?* – may not yield a binary answer, and so require a careful weighing of how the data and algorithms balance in the flesh and blood world.

One unambiguous way that AI supports autonomy is by catalyzing experimentation. Part of the essence of human freedom is the ability to try new things, and AI contributes when helping users access possibilities – professional, romantic, cultural, intellectual – outside the funnel of those already established by their habits. The challenge is to mechanically produce serendipity.

The challenge is significant, as anyone has learned who has scrolled Netflix hoping for a movie that is enjoyable but also unfamiliar to established tastes. Part of the problem is the way Netflix predicts satisfying recommendations. In a recent public talk, a company engineer spent nearly his entire period describing techniques to isolate new films that significantly *matched* those specific viewers had already liked. The strategies include extrapolating from what the client has already seen, to finding other viewers who resemble the target subject, and checking what *they* like. Not until the final sentences of the presentation did he reveal an effort within Netflix to help viewers escape the logic of similarity and discover unexpected possibilities:

> Production biases in ML models can cause feedback loops to be reinforced by the recommendation system. We do research and development in the causal recommendation space specifically to get our recommender models out of this feedback loop (Deoras 2020).

It was a tantalizing conclusion, but also frustrating as no details were provided.

For autonomy scoring, those last sentences merit follow-up. *Can Netflix provide users with successful film suggestions that their users could not have foreseen wanting, that the user may have skipped over if left to their own devices?* If so, the machine is bettering human recommendation. It is also creating new opportunities, expanding self-determination.

Technically, AI serendipity means helping users escape the trap of their own accumulated data. Statistical work addressing the challenge is currently underway in the area of social media polarization (Celis 2019: 160). Online users reliably maintain interactions with others who share their beliefs and values, leading to an echo-chamber: views feedback and intensify in a shrinking circle (Bozdag and van den Hoven 2015). One serendipity response constrains selection algorithms to contain examples from imperfectly related ("non-optimal") groups (Celis 2019: 168). Possibly, this intentional error is a step toward mechanically provoking serendipity. More work will need to be done conceptually and technically, but whether it is Netflix viewers seeking unexpectedly delighting movies, or social media participants seeking unfamiliar but provocative connections, the autonomy tension is the same. Big data and predictive analytics can reinforce the conveniences and pleasures of the familiar, or they can diverge. Divergence serves autonomy.

There is also the money question. On one side, tightening personalization in AI provisioning of user opportunities is pervasive online because the efficiency presumably creates higher revenue for the platform (Sakulkar and Krishnamachari 2016). Still, boredom is a human reality, and some research indicates that providing fruitful discovery opportunities for users holds their attention better than simple repetition of what has already proved satisfying (Kamehkhosh *et al*. 2020). Serendipity, in other words, may dovetail with conventional business incentives.

Summarizing, autonomy as a key performance indicator for AI ethical investing measures whether the AI oppresses or vitalizes self-determination.

*Figure 2*

| AI Ethics of Personal Freedom | |
|---|---|
| **Autonomy:**<br>***Does the AI***<br>***debilitate***<br>***or invigorate***<br>***self-determination?*** | Negative<br>• Platforms diminish opportunities<br>• AI decides *for* users<br>• Habituation and dependence fostered<br><br>Positive<br>• Platforms open opportunities<br>• AI enables better user decisions<br>• Opportunities for experimentation created |

**Dignity**, the second personal freedom criterion, requires that people be treated as ends in themselves, and not only as means to someone else's ends (Kant 1996: 429). The dignified hold intrinsic value, as opposed to tools which hold value only instrumentally, in order to do something else.

Human dignity begins as freedom from exploitation, which precludes being understood purely as data for processing, profit-taking, and deleting because the dignified have their own independent projects intrinsically meriting recognition. Then, to complete that independence, dignity is also freedom from patronization, which means that taking responsibility for one's own acts is not a burden but a positive right. The terminal example of this right to accountability is the execution of murderers. Dignity recommends it not as an obligation to the original victim, or to society, or as a way of dissuading future crime, instead, it is an expression of *respect for the murderer* (Kant 1996: 333). Failure to exact the ultimate penalty is not merciful or benevolent but insulting: the criminal granted clemency is treated like a child, belittled as incapable of making decisions and holding responsibility for them.

Like autonomy, the AI contribution to human dignity is difficult to abstract from corporate disclosures and public sources. Nevertheless, careful analysis can distinguish AI companies that treat users as ends in themselves, from those treating users as means to the *company's* ends.

In 2014, a co-founder of the OkCupid dating site published a blog entry titled *We Experiment on Human Beings!* It has since been removed, but while visible at least some of the experiments were recounted (Berinato 2014). In one, users who the OkCupid algorithms determined to be incompatible were told the opposite. When they connected, their interaction was charted by the platform's standard metrics: how many times did they message each other, with how many words, over how long a period, and so on. Then their relationship success was compared against pairs who were judged truly compatible. The test presumably measured the power of positive suggestion: Do incompatible users who are *told* they are compatible relate with the same success as true compatibles? (Rudder 2014)

The answer is not as interesting as the users' responses. One asked, "What if the manipulation is used for what you believe does good for the person?"

(Rudder 2014) The appeal here is to the fine print of the dignity requirement: treat others as ends and not *only* as means. In the real world, it can be true that exploiting others mixes with helping them. The question dignity asks is: Which one serves the other? Are the romance-seekers being manipulated in experiments for *their* ultimate benefit because the learnings will result in a better platform and higher likelihood of romance? Or, is the experiment more about the platform's owners and their marginally perverse curiosities?

Part of the answer lies in the fact that the experiment – and therefore the manipulation – was revealed to the users, leaving them free to respond. It is not clear how many responded by cancelling their accounts (Berinato 2014). In any case, the dignity question for AI is not whether the technology helps users, it is whether users *determine for themselves* what the word "help" means.

The other side of dignity is freedom from condescension. This can be a stern demand. AI chatbots, for example, are increasingly employed to ward off depression, especially among the elderly (Pereira and Díaz 2019). The chatbots are also increasingly difficult to detect as mechanical instead of human (Shestak *et al*. 2020). Further, patients respond better to interlocuters who they believe to be human (Chan *et al*. 2017). Those premises write their own conclusion: deceptive AI chatbots should be deployed to elderly patients. The result would likely be diminished depression, but as long as the patients are not informed of the AI impersonation it remains true that the entire process depends upon patronization. The dignity objection is that users are being treated as unworthy of fending for themselves when it comes to their own treatment. So, the problem with deceitful AI is not exactly that decisions are being made *for* patients (that is the autonomy objection), it is the implication that patients cannot manage the deciding, they are unworthy of accountability for the decision. When that happens, the dignity score must be adverse.

As an ethical performance indicator, dignity measures whether an AI respects users' independent projects, and respects users taking responsibility alone for where those projects lead.

*Figure 3*

| AI Ethics of Personal Freedom | |
|---|---|
| **Dignity:**<br>***Does the AI respect users' independent projects, and respect users taking responsibility alone for where those projects lead?*** | Negative<br><ul><li>Users as objects, tools, means to others' ends</li><li>Users not treated as responsible for actions</li></ul>Positive<br><ul><li>Users as subjects with intrinsically valuable projects, ends-in-selves</li><li>Users alone responsible for their data and algorithm decisions</li></ul> |

**Privacy** is the third personal freedom metric, and defined as control over access to our own personal information (Westin 1968: 3). Because privacy is an ability, not a state, it cannot be measured by how many people know how much about someone, instead, privacy gauges *the power to determine* who knows how much. Kim Kardashian, for example, is one of the most private people in the world, which does not mean her personal life is closely kept, but it *is* closely guarded: she strictly controls her own exposure. The fact that she chooses overexposure by conventional standards subtracts nothing from her privacy.

Several years ago, a news report circulated of a woman who lived nocturnally as a sex worker, while maintaining an ordinary daytime identity with an academic email address and typical social media postings. The two worlds kept their distance, until she and her clients began appearing in each other's "People You May Know" recommendations on Facebook (Hill 2017). She tried to stop the connections, but was overpowered by the data and algorithms, and so learned first-hand the difference between degree of intimate availability which has nothing to do with privacy, and *control* over that availability which is privacy.

The reason for privacy – and the reason it exists as a category of personal freedom – is to decide for ourselves who we want to be. Normally, the decision does not widen to professor or prostitute, but in smaller ways all of us depend on limiting personal information to form definitions of ourselves as we go through a typical day. Parents display a goofiness in front of their young children that they would be appalled to reveal to their coworkers. The persona many adopt in the workplace would aggravate a spouse, and spouses define each other in unique ways when no one is watching. So, control over access to personal information is not an occasional concern, it is an everyday part of creating an identity: at any given time and in the company of selected others, we craft our own identity by exposing parts of ourselves, and equally by concealing others. By exercising privacy, we become who we are.

On the practical level of scoring for responsible investing, significant advances have been made. The General Data Protection Regulation (GDPR) is a milestone advance, and departing from Article 5 – *Principles Relating to Processing of Personal Data* – researchers in Swedish and Danish universities have assembled specific criteria in the areas of data governance and cyber security that can measure privacy performance. With slight modifications, they are (Vinuesa et al. 2020b):

- Users control their own data's collection and use.
- Data collected and used transparently.
- Data-minimization principle in effect: only information necessary to perform the AI function is gathered, data storage is local, and temporary.
- Privacy-by-design engineering.
- Security ensured by user authentication to prevent risks such as access, modification, or disclosure of data.
- A cybersecurity yield is available, one that measures the magnitude and efficiency of a company's security expenditure in relation to the value at risk (Nolan *et al*. 2019).

Within today's investor research community, privacy is an area where the analysis of AI-intensive companies overlaps with traditional ESG reporting. The category, consequently, is well-established. Sustainalytics's *Managing data privacy risk: comparing the FAANG+ stocks* assesses how seven major technology corporations perform in data privacy. Facebook and Amazon are graded as vulnerable to high-risk exposure attributable to weak data management. Apple is reported to be well-positioned due to strong data governing policies (Sustainalytics 2018). Another sector leader, MSCI, benchmarks more than 600 companies annually on risk linked to privacy through one of their 37 ESG Key Issues: *Privacy and Data Security* (MSCI 2019). The larger result is that the information required to efficiently score AI-intensive companies on their privacy performance is increasingly available.

Summarizing, privacy as an ethical investment performance indicator measures whether an AI adds to, or subtracts from users' control over access to their own personal information.

*Figure 4*

| AI Ethics of Personal Freedom | |
|---|---|
| **Privacy:**<br>***Does an AI add to, or subtract from users' control over access to their own personal information?*** | Negative<br>• Personal data collected indiscriminately<br>• Opacity about where personal information goes, how used<br>• Cybersecurity inadequacy<br><br>Positive<br>• User controls personal exposure<br>• Transparent data collection and use<br>• Data-minimization principle for collection, use, storage<br>• High cybersecurity |

## Social Wellbeing Metrics: Fairness, Solidarity, Sustainability

**Fairness** is traditionally defined as equals treated equally, and unequals proportionately unequally (Aristotle 1934: Book 5:3:13). If two people have similar financial backgrounds and apply for comparable loans, then data and algorithms combined to produce lending decisions should arrive at similar disbursement results. Conversely, to the extent two applicants present unequal financial strength (different income levels, outstanding debts, and similar), their loan terms should be proportionately dissimilar. For AI designers within this lending model, fairness is straightforward. Information is processed to predict who will – and who will not – repay a loan, and the resulting probability corresponds with a loan application decision entailing acceptance or rejection, as well as an interest rate. Being fair, consequently, means predicting accurately. Ideally, those applicants who will repay do receive loans, and those who will not, do not.

Fairness can also be understood in terms of identity groups. Frequently advocated under the title of social justice, the idea here is not that two people who are similar in terms of their finances are also determined to receive similar credit opportunities, instead, balances are sought between men as a group and women grouped, or between races, or other communities. To explore these fairness concepts, several statistical models have been developed. Equalities in algorithmic lending can be sought, for example, as the fraction of non-defaulting members from racial groups (Hardt *et al*. 2016: 17-19; Narayanan 2018: 00:34:06). In other words, if you take all those individuals who were awarded loans, and divide them into race categories, and then check the proportion of members in each category that repaid, the percentages should be about equal. If they are not, if one racial aggregate contains relatively few defaulting members, that suggests mediocre credit risk applicants in that group are getting rejected, while in other groups mediocre risks are getting accepted as reflected by their relatively high default proportion. So, the ideal of equal opportunity between races may justify reweighing the lending algorithm to bring nonpayment proportions into alignment.

The overall result is a distinction between two fairness views: one is about individuals and treating them equally, while the other concerns groups and treating them analogously. In technical terms, the debate is between calibration (individual accuracy) and parity (group balance), and the fairness dissonance between these two possibilities has been among the highest profile discussions in recent AI ethics, notably involving the Northpointe technology company and its model for predicting recidivism risk (Washington 2018; Angwin *et al.* 2016).

For the purposes of AI ethical investing, there is no right or wrong here. What counts is awareness that and how the sides exist: when AI designers opt for calibration, or for parity, they are making a fairness decision that should be justifiable, or at least explicable. An AI intensive company with a strong fairness score is one that knows where it stands, and why.

Fairness is not only about distribution of loans and other opportunities, it is also about *representation*. Google searching "CEO" images turns up overwhelmingly white male faces, which corresponds with the gender and race reality (Lam *et al.* 2018). The representational fairness question emerges when that truth crosses this one: people may be less likely to aspire to be a CEO if they do not perceive others like themselves have already followed that route (Barocas and Selbst 2016).  Here again there rises the dichotomy between fairness resting on accuracy (CEO image search should produce truly representative CEO images), and fairness resting on opportunity (CEO image search should be tweaked to invite all of society into the aspiration). This is a true dilemma – accuracy or opportunity – but for the scoring of AI ethics, again, no resolution is required, only awareness that algorithmic weightings are fairness decisions *and* decisions about what counts as fairness.

A third and more straightforward AI intersection with fairness involves the data used for training applications. It may reflect individual, cultural, or historical biases (Gianfrancesco *et al.* 2018, Char *et al.* 2018) which may lead to unwarranted disadvantages in treatment for individuals or groups (Bobrowski and Joshi 2019, Goodman *et al.* 2018). At Amazon in 2015, a workforce where male employees were superior to their female counterparts in terms of quantity infected data employed to algorithmically rate job applications with the bias that male employees were superior in terms of quality. Female resumes were correspondingly downgraded (specifically those featuring graduation from two women's colleges, as well as various other word combinations, including "women's chess club") (Kodiyan 2019: 2). The AI was rewritten, and then discontinued, but the risk persists that algorithms can start with uneven information, recycle it, and repeat as the outcome tilts ever further out of balance. Part of AI fairness, consequently, involves data inspection: where did the original information come from? How might non-material factors pollute it? Because equal outcomes depend on balanced initial information (IEEE 2019:190), an AI ethics score partially reflects a company's attentiveness to the quality of its data.

Finally, the IEEE Global Initiative on Ethics of Autonomous and Intelligent Systems adds that a diverse workplace may help AI designers encounter and remedy bias problems, both in the initial data and in subsequent iterative processing. Diversity in this context refers not only to legally stipulated identity groups, but also to educational and professional backgrounds: interdisciplinary teams might include computer scientists along with experts in medicine, architecture, law, philosophy, psychology, and cognitive science (IEEE 2019: 126, 19). To the extent this is correct, workplace diversity itself may be gauged as a proxy for protection against the risk of unfairness.

Condensing the discussion, fairness is scored as equal treatment for society's members. Key elements of the performance indicator include:
- AI designers are knowledgeable with respect to the accuracy versus opportunity debate at the core of AI fairness.
- Safeguards are erected against biased (in social and statistical senses) training data.
- Workplace diversity may serve as a proxy for protection against the risk of unfairness.

*Figure 5*

| AI Ethics of Social Wellbeing | |
| --- | --- |
| **Fairness:**<br>*Are all*<br>*treated equally?* | Negative<br>• No engagement in accuracy versus opportunity debate<br>• Failure to account for bias embedded in data, or exacerbated in processes<br>• Minimal workplace diversity<br><br>Positive<br>• Engagement in accuracy versus opportunity debate<br>• Safeguards against bias embedded in data and exacerbated in processes<br>• Genuine workplace diversity |

**Solidarity** is the inclusiveness of no one left behind (Microsoft 2018). In AI medicine, because the biology of genders and races differ, a diagnostic or treatment may function well for some groups while failing others (Noor 2020; Wang and Keng 2018). The ethical difficulty is captured by a hypothetical: If an AI scan analysis for Melanoma is trained on data from white men, and it proves effective, should white males who may have the disease wait to use the technology until data has been gathered and training administered for both genders and all races? A strict solidarity posture could respond affirmatively, while a flexible solidarity would allow use to begin so long as data gathering for unrepresented groups also initiated. Solidarity's absence would be indicated by neglect of potential users, possibly because a cost/benefit analysis returns a negative result, meaning some people get left behind because it is not worth the expense of training the machine for their demographic segment.

Another solidarity element is a Max/Min distribution, one where the benefits of an AI are distributed maximally to those who have least (Rawls 1971: 266). In some cases, it does seem likely that AI tilts advantages toward the disadvantaged. Psychological carebots designed to help fight depression may be more affordable and accessible than human, face-to-face treatment, which implies that their development provides a mental health advantage to some who previously could not afford it (Singh 2019). By contrast, there are indications that AI as an industry is exacerbating inequalities instead of remedying them. According to one thorough study, "Artificial Intelligence puts more low-skilled jobs at risk than previous waves of technological progress" (Nedelkoska and Quintini 2018). It may result, of course, that job losses in one sector create better or new opportunities elsewhere. Regardless, on the microeconomic scale and with respect to rating specific AI products and companies in terms of Max/Min, a positive score goes to those bringing the greatest benefits to users who have the least. More broadly, a positive score is assigned when a technology is optimized to leave no one behind.

*Figure 6*

| AI Ethics of Social Wellbeing | |
| --- | --- |
| **Solidarity:** *Is no one left behind, with the most going to those who have least?* | Negative: <br> • Access to AI is exclusive <br> • Benefits clustered among the most advantaged <br><br> Positive <br> • Access to AI is inclusive <br> • Most benefits distributed to the least advantaged |

**Sustainability** means ending poverty and hunger, protecting the planet from environmental degradation, and fostering peaceful, inclusive societies according to the *2030 Agenda for Sustainable Development* (Assembly 2015: 5). These are a dauntingly broad ambitions, though they can be grounded in the United Nations 17 Sustainable Development Goals (SDGs) initially published in 2015 (United Nations 2015). These silos are convenient for evaluating AI-intensive companies primarily because they are well known to ESG researchers, and already feature prominently in their investing research, strategy, and publishing (MSCI 2019b).

One representative publication reports the following:

> In SDG 1 (No poverty), SDG 4 (Quality education), SDG 6 (Clean
> water and sanitation), SDG 7 (Affordable and clean energy), and SDG
> 11 (Sustainable cities), AI may act as an enabler for all the targets by
> supporting the provisioning of food, health, water, and energy services
> to the population. AI can enable smart and low-carbon cities
> encompassing a range of interconnected technologies such as electrical
> autonomous vehicles and smart appliances that can enable demand
> response in the electricity sector with benefits across SDGs 7, 11, and
> 13 on climate action (Vinuesa et al. 2020: 2).

Technology can also diminish sustainability:

> AI may also lead to additional qualification requirements for any job,
> consequently increasing the inherent inequalities and acting as an
> inhibitor towards the achievement of this target (Vinuesa et al. 2020:
> 2-3).

The next step is to repeat the analysis, but no longer as applied to the
industry and instead as aimed toward particular AI-intensive companies.
For example, the firm AgrilogicAI (later rebranded as Dagan Tech) uses
machine learning methods to identify high-yielding soybean variants by
analyzing data from remote sensing and soil features. According to their
report:

> Collectively, our models identified fifteen elite varieties from 21
> predictive variables to forecast soybean yields in 2015 at 58 test
> locations. This method can boost commercial soy yields by about 5%
> and shorten the time for commercial variant development (Aviv 2018).

So, the technology helps maximize yield for specific soil conditions, and
speeds crop optimization, which serves SDGs 1 and 2 (Poverty, Hunger), as
well as 15, which seeks to protect, restore and promote sustainable use of
terrestrial ecosystems. Similar points could be made about the company's
*Farm360AI* platform which predicts corn and soybean yields from satellite
imagery and weather data. Overall, the enterprise's reported innovations
suggest a high score in the sustainability metric.

Shifting to practical concerns, one significant benefit of aligning AI sustainability with United Nations' criteria is that data corresponding with major companies' performance – as well as ratings providers' results – are abundant, and already available to investors. Pricewaterhouse Coopers's 2019 SDG Challenge investigates over 1,000 company reports on their engagement with sustainability (Scott and McGill 2019). In 2020, S&P Global analyzed 150 categories aligned with the SDGs across 3,500 companies representing 85% of global market capitalization (Trucost 2020). Bloomberg reports that at least a dozen major third-party enterprises provide independent ratings of companies' SDG sustainability performance, including Sustainalytics, MSCI, Moody's, and Fitch Ratings (Poh 2019). More, robust investigation and research by nonprofit organizations adds a further layer of accessible resources for financial analysts. The new AI for SDGs Center, for example, is dedicated to detecting and scaling up use cases for AI enabling the 17 SDGs (Miailhe *et al*. 2020).

In sum, a sustainability score reflects a contribution to social wellbeing as gauged by the United Nation's Sustainable Development Goals.

*Figure 7*

| AI Ethics of Social Wellbeing | |
|---|---|
| **Sustainability:** *Does the AI promote enduring social wellbeing?* | Negative <br> • Slows progress toward the 17 UN Sustainable Development Goals <br><br> Positive <br> • Speeds progress toward the 17 UN Sustainable Development Goals |

## Technical Trustworthiness Metrics: Performance, Safety, Accountability

**Performance** measures how well a machine works, it is the accuracy and efficiency of AI outputs. In human experience, the question about functional success can often be answered in terms of personalized convenience.

At a 2020 professional AI conference, a Netflix machine learning research scientist was asked, "How is Netflix using AI for a positive impact?" He responded:

> We try to build models for recommendations that maximize Netflix members' enjoyment of the selected item while minimizing the time taken to find it. Enjoyment integrated over time i.e. goodness of the item and the length of view, interaction cost integrated over time i.e. time it takes the member to find something to play, are some of the factors we consider while building our ML/AI models for a positive impact on our 100M+ members (Deoras 2020).

For distinct AI companies and operations, the meaning of personalized quality will shift, but the Netflix metrics – enjoyment integrated over time, interaction cost integrated over time – double as objective ethical scores. They fluctuate uniformly, and a machine attaining engineering goals simultaneously fulfills a humanist principle. It follows that performance is, in a limited sense, the easiest investing category: the better the AI is technically, the better it is ethically.

Another way to score performance is in relative terms. If you could use only Google, or only Bing for a year, which would you choose? Market share may provide a simple and revealing answer to the question about measuring accuracy and efficiency.

A similarly relative evaluation could be performed with AI set against human providers. In 2019 the computer scientist Geoffrey Hinton tweeted a widely circulated thought experiment:

> Suppose you have cancer and you have to choose between a black box AI surgeon that cannot explain how it works but has a 90% cure rate and a human surgeon with an 80% cure rate? (Hinton 2020)

The answer – perhaps as provided by a focus group or, more pointedly, by actual patients in a hospital – may produce a useful and double evaluation of AI individualized service as it relates to the larger principle of technical trustworthiness. First, the raw numbers could be tested: Does the AI really out-cure the human surgeon? Second, and assuming the curative percentages are confirmed, how large must the outperformance be for patients to opt for the AI over the security of a human voice and hand?

It is now documented that in certain medical fields AI *does* outperform humans, and not just in diagnosis but also in what can be processed on the way to diagnosing (Grote and Berens 2019). Cancer screenings, for example, test human doctors in two ways: sensitivity to anomalies (how well they see), and rapidity of scan readings (how much they see). The latter is increasingly significant because new cancer detection technology seeks to increase sensitivity by multiplying images (Conant *et al*. 2019). Theoretically, there is no upper limit, millions of scans could be sliced from any one patient. For human doctors, however, most would be superfluous since there are not enough hours in the day to examine them all. AI does not have that problem: the machine could potentially scan the images as rapidly as they are produced. It follows that a true performance rating will not only account for how well a task is accomplished (image analysis), but what tasks become possible (rapidly analyzing streams of images) when AI is doing the performing.

Finally, a critical aspect of performance as an ethical criterion is its power to render *other* ethical criteria moot. The safety of an AI-guided process, for example, is a significant human concern, but only when performance is sub-optimal. If the data and algorithms controlling a driverless car *never* fail to take the best possible action, then the issue of vehicular safety will never arise. Practically, that ideal may remain unattainable, but it remains true that an exemplary performance score significantly diminishes safety worries. The same point could be made about accountability. Vast resources currently pour into projects to ensure that AI decisions in healthcare, loan disbursement, employment opportunities and similar are transparent for their users so that responsibility may be assigned in case of error: to have accountability, it is first necessary to define exactly what went wrong (Tjoa 2020). That need evaporates, however, to the extent that errors dissipate. At the extreme – when there are no complaints – it does not matter who is at fault. Performance, therefore, is an ethical good not only intrinsically but also instrumentally, as a way to assuage other ethical concerns.

*Figure 8*

| AI Ethics of Technical Trustworthiness | |
| --- | --- |
| **Performance:**<br>***How well does the machine work?*** | Negative<br>• Accuracy, efficiency, personalization, convenience inferior to competitors in category, or human-based alternatives.<br><br>Positive:<br>• Accuracy, efficiency, personalization, convenience superior to competitors in category *and* human-based alternatives.<br>• Performance diminishes other threats to technical trustworthiness. |

**Safety** as a criterion of technical trustworthiness asks whether an AI is dangerous.

In 2016, a Tesla crashed into a truck. According to Tesla:

> The Model S was on a divided highway with Autopilot engaged when a tractor trailer drove across the highway perpendicular to the car. Neither Autopilot nor the driver noticed the white side of the tractor trailer against a brightly lit sky, so the brake was not applied. The high ride height of the trailer combined with its positioning across the road and the extremely rare circumstances of the impact caused the Model S to pass under the trailer, with the bottom of the trailer impacting the windshield of the Model S. (Tesla Team 2016.)

This horror movie accident represents a particular AI fear: a machine capable of calculating pi to 31 trillion digits (Porter 2019) cannot figure out to stop when a truck crosses in front. The power juxtaposed with the debility seems ludicrous, as though the machine completely lacks common sense which, in fact, is precisely what it does lack (De Freitas 2020: 8). For human users, one grievous effect of the debility is no safe moment. As with any mechanism, AIs come with knowable risks, but it is beyond that, in the region of unknowable

perils – especially those seemingly easily avoidable, even for children – that human trust in AI destabilizes.

This is the deep problem: machines and humans create knowledge differently. AI filters for correspondences, while humans impose linear sequence and causality onto raw perceptions (Kant 1997: A91/B124). This difference – correspondence versus causality – at the origin of knowledge itself means that machine learning *cannot* be understood, not even by infinitely rapid human thinking. The true machine-human divergence, in other words, does not concern velocity or power of reasoning, instead, it is about what the verb *to reason* means, and that renders everything coming afterward irreconcilable. So, the way AI produces knowledge is inhuman, which does not falsify the knowledge, but it does preclude comprehensive human knowledge about the knowledge.

Because decisions guided by correspondence will always be prone to catastrophes that are by nature inconceivable for those who suffer, scoring an AI company for safety becomes disorienting. It is not just that perfect confidence is impossible, but also that there is no way to delimit the potential scenes of danger: there is no way to even begin to comprehend what might possibly go wrong. Consequently, safety can only be conceived as a process instead of a goal. Improvements may be marked by remedying encountered problems like the Tesla failing to detect the truck, but since the remaining threat cannot be calculated, there is no way to know that an ultimate safety goal has been reached, or even how close we are to it.

In their paper *A Safety Standard Approach for Fully Autonomous Vehicles* the authors write that, "it is important to address known safety issues before exposing testers and the public to undue safety risk," and add:

> Rather than adopting a fiction that mere conformance to a standard at deployment results in flawless risk mitigation, it is important to continually evaluate and improve the residual risk present in the system. Honest self-assessment and iteration over the system development and deployment lifecycle is vitally important to mature the safety case (Koopman *et al*. 2019: 6).

This is a roundabout way of saying that we should make driverless cars as safe as reasonably possible at the start, and then when accidents occur, learn why as best we can, and then redesign the AI to avoid recurrences. Users, that means are irretrievably crash test dummies.

Another response to safety challenges is human oversight. A designer monitoring the AI ("Human on the loop") or a deployment supervisor accompanying the AI ("Human in control") is granted the power – and the responsibility – to adjust the machine's actions. Tesla tapped into this AI defense strategy when responding to the truck crash. After explaining what happened out on the road, and expressing condolences, the company curtly added:

> When drivers activate Autopilot, the acknowledgment box explains that Autopilot "is an assist feature that requires you to keep your hands on the steering wheel at all times," and that "you need to maintain control and responsibility for your vehicle" while using it (Tesla Team. 2016).

The message is that human users are empowered to override AI decisions and so shave off at least those dangers obvious to us and invisible to machines. On the other hand, if users need to be driving along, what is the point of Autopilot?

Ultimately, safety as an AI Human Impact performance indicator needs to be rendered calculable and meaningful. The E.C. *Ethics Guidelines to Trustworthy AI* lists considerations that analysts could convert into scoreable categories. Beyond the intrinsic threat of the technology's unpredictability, the document adds external dangers to AI dependability. They may arise intentionally due to bad actors, or unintentionally due to misuse of an AI system, and safeguarding against them counts in favor for a safety score. The silos include:

- Fallback systems that ask for human operation in the face of irresolvable problems.
- Protections against hacking.
- Accounting for unintended uses (AIHLEG: 16-17).

Another measuring possibility uses resource allocation as a safety proxy: the more money and expertise a company dedicates to ensuring its AI mistakes are rapidly and well corrected, the higher its score. Of course, the importance of safety itself depends on the magnitude of risk posed by a system's capabilities (AIHLEG: 17). Driverless cars and autonomous floor cleaners present distinct dangers and require different investments to qualify as safe.

Empirical safety results may also be measured either across the industry, or comparatively between AIs and humans. A safety score for Tesla may be initially calculated by weighing deaths per mile driven (or collisions per mile driven, or a similar anomaly) against those attributable to other autonomous vehicle companies. Next, the relativist strategy could be applied between Tesla and human drivers. Either way, there are no guarantees. There are not even confident probabilities of safety since we can never know the full extent of the peril.

*Figure 9*

| AI Ethics of Technical Trustworthiness | |
|---|---|
| **Safety:**<br>*Is the AI dangerous?* | Negative:<br>• Unreliable fallback mechanisms<br>• Inadequate protections against hacking and unintended uses<br>• Resources dedicated to safety incommensurate with risk<br>• AI statistically less safe than competitors in category *or* human-based alternatives<br><br>Positive:<br>• Efficient fallback mechanisms<br>• Robust protections against hacking and unintended uses<br>• Resources dedicated to safety commensurate with risk<br>• AI statistically safer than competitors in category *and* human-based alternatives |

**Accountability** measures how well responsibility is assigned for AI-determined actions.

One way to train an autonomous vehicle is through demonstration. A human driver takes the lead by operating a camera-outfitted car, and as kilometers and data accumulate, the AI is increasingly able to make decisions by imitating those observed on the road (Kebria 2020). Potentially, human owners could train their vehicles with traits of their own driving: distance between cars while cruising, acceleration rate, breaking abruptness, turning radiuses, all that could be personalized. Later, and with autopilot engaged, the car crashes. Who is to blame? The car owner? The AI? The AI designer?

Responding starts with explainability, with locating and describing the factors contributing to the machine-driven accident. Returning to the Tesla fatality, a report from the European Parliament found that the car's computer mistook the white tractor trailer crossing in front for clouds in the blue sky, and did not even slow down (Panel 2020: 34). That made for a frightful accident, but also the beginning of a comprehensible AI event: because it was possible to broadly grasp the algorithmic decision in human terms, an investigation into responsibility became possible.

Narrowing the explanation can be challenging. Defining *why* the truck was classified as sky entails grasping how the AI weighted and combined incoming data. There are sophisticated mathematical approaches to this kind of problem developed from game theory (Das and Rad 2020: 8). In some cases, though, human understanding may not fully capture the machine's algorithmic methods (Zerilli 2019: 670) and a lesser standard may be applied, possibly interpretability (Gall 2018) which is about predicting what will happen instead of why. It is foreseeing what output will follow from an input, as opposed to following along to determine exactly how the inputted data is processed to generate a result. Regardless, the pursuit of explainability and its limits is a rich area of contemporary AI research (Das and Rad 2020), and analysts who are assigning an accountability score will weigh the resources a company dedicates to understanding how its technology functions as it does.

A hotly debated tradeoff in this region of accountability weighs explainability against AI performance (Whittlestone et al 2019: 20). If humans cannot keep up with AI processing, and if explainability is a priority (as it may be in some areas of critical healthcare), then the machine can be slowed down. Gaussian processes, neural networks and random forests can be replaced by linear regression or a single decision tree. These kinds of changes represent, in effect, an exchange of computing power for a better view of the computations. The choice behind it – setting a balance between knowing, and knowing why there is knowing – may be answered one way or the other, but in terms of ethical investing, a positive explainability score may derive from the willingness to sacrifice at least some performance.

More significantly, there is a true dilemma here for AI investors. Ideally, accountability and performance are *both* maximized. In practice, though, that may not be possible, meaning a choice gets forced. If it is, and if the choice is between two positive values, then the strategy of AI Human Impact provides no guidance. The neutrality is even by design because the rating's guiding purpose is to increase investor freedom by facilitating informed decisions. Rendering the decision clearly in terms of the criteria set into tension – performance versus accountability – *is* the positive contribution of the rating strategy because it helps investors understand what is at stake ethically before making their own decisions about money and technology.

Accountability begins with explainability and culminates with *redress*, with the ability to respond to – or recover compensation for – an AI function after an error has been located. Legal studies influence significantly here, including the doctrine of the learned intermediary (Harned *et al*. 2019) which generally holds in the medical field that when an instrument causes harm, the human operator is responsible. If an AI cancer diagnosis proves erroneous, it is the doctor who signed the finding, not the machine that would be the target of a lawsuit (Sullivan and Schweikart 2019).

There is a structural problem, however. A central AI benefit is an increasing ability to function without human oversight: the reason Tesla develops autopiloting cars is not so that drivers can ride along gripping the wheel, it is to allow a nap on the way home from work. So, the further a machine learning platform advances, the more it may extend from any learned intermediary, and that means the better the AI, the harder it is to pin blame on a human overseer.

One solution is to blame the AI directly by assigning legal personhood to it, as is done for corporations (Hildebrandt 2019).

Another solution is redress-by-design (Quintarelli 2019), which is the engineering strategy of formulating AIs so that harms can be identified and corrected rapidly, and also so that outputs can be *contested* (Ploug and Holm 2020; GDPR: Article 22). For example, big data and algorithms are increasingly employed to make lending decisions because loan distribution can be reduced to predictive analytics estimating the risk of default (Verma and Rubin 2018). When a loan is denied, redress-by-design may help applicants understand what specific piece of data led to their rejection, and enable the opportunity to object effectively.

More broadly, the E.C. *Guidelines* establishes elements for adequate redress in AI systems (AIHLEG 2019: 31). There should be:

- a way for users to contest decisions
- a way for users to make a claim for harm done
- available information about how to make the claim
- available information about the circumstances that may occasion such a claim

In the end, accountability as an ethical investment performance indicator measures how well responsibility can be assigned for what an AI does. It encompasses – and can be scored as – explainability and redress.

*Figure 10*

| AI Ethics of Technical Trustworthiness | |
| --- | --- |
| **Accountability:** *How well can responsibility be assigned for what an AI does?* | Negative<br>• Limited explainability; the AI is a blackbox<br>• Few or obscure mechanisms for contestation and redress<br><br>Positive<br>• The specific data most responsible for an AI output is identifiable<br>• Ample contestation and redress opportunities |

## How are the Categories Scored?

Following Vinuesa (*et al*. 2020), a three-point metric may be employed to score ethical performance indicators in AI intensive companies. A score of 2 corresponds with a positive evaluation, 1 corresponds with neutral or not material, and 0 corresponds with inadequacy. The scores convert into objective investment guidance, both individually and as a summed total.

Investors who are particularly interested in privacy, for example, or safety, may choose to highlight those metrics in their analysis of investment opportunities. Others may widen the humanist vision to include the full range of AI ethics concerns, and so focus on the overall impact score derived from a company or a specific technology.

The scoring rubric is itemized below. Because personal freedom is the orienting metric of responsible AI investing, it is double weighted in the aggregating formula.

- Personal Freedom
    - Autonomy (Score 0 - 2)
    - Dignity (Score 0 - 2)
    - Privacy (Score 0 - 2)

- Social Wellbeing
    - Fairness (Score 0 - 2)
    - Solidarity (Score 0 - 2)
    - Sustainability (Score 0 - 2)

- Technical Trustworthiness
    - Performance (Score 0 - 2)
    - Safety (Score 0 - 2)
    - Accountability (Score 0 - 2)

- AI Human Impact Score
    - Total PF*2 + SW + TT = (0 – 24)  / 2.4 = Net Score on 10 scale

Additionally, the categories may be represented as tensions. AI employed to treat elderly depression sufferers through chatbots masquerading as humans creates a tension between human dignity on one side, and on the other sustainability particularized as Sustainable Development Goal 3, Health and Wellbeing at all ages. Similarly, the ideal of a high privacy score pulls against performance across a range of AI applications, from product recommendations to facial recognition systems screening for violent criminals. Performance can also prove irreconcilable with accountability when an effective AI is also an inscrutable blackbox. In every case, what is significant is not that one or the other direction is preferable, there is no transcendent reason why a high score in one category must be preferred to another. Instead, the significance lies in delineating ethical commitments bound up with any preference. That clarity allows individuals and groups to align their finances with *their own* values. AI Human Impact fabricates investor freedom.

## Limitations and Further Research

The most immediate limitation of this essay is that more specificity will be required to apply AI Human Impact in any particular case, and the process will be laborious though manageable, as is illustrated by a multidisciplinary group that includes this paper's author, and that works out of the Frankfurt Big Data Lab in Germany. Though not oriented toward investors, the academic group evaluates startup AI companies in ethical, legal, and technical terms, and the group's publications provide an indication of the workflow implied by AI Human Impact (Zicari *et al* 2021).

At least two other directions for further research are apparent. First, there remains an open question about what counts as an AI-intensive company. A robotics-intensive startup, for example, may expand into AI development to complement its core mechanical product. At what point does an ethical evaluation become recommendable? To answer, the elaboration of a technically sophisticated definition of an AI-intensive company would be helpful.

At the other end of the spectrum, there is a question about how the model works when an expansive company employs AI in multiple ways and forms across its operations. Amazon, for example, has applied machine learning to internal employment decisions, product recommendations, delivery logistics, and more, and the consequent tangles for responsible investors have proven considerable (Beslik 2020). To face the complexity and organize a useful response, the ESG model could provide useful guidance as an analogous problem is faced when a single sprawling company interacts with environmental, social and governance issues across a diverse range of contexts (Chen *et al*. 2017, Drempetic *et al*. 2020).

## Conclusion

ESG investing is commanding increasing assets (GSIA 2018) while AI-powered companies are consuming a larger share of the global economy (Vinuesa 2020). Both are growing, but they are also growing apart. ESG was forged amid conflicts over exploiting natural resources and competition between social collectives (MSCI 2020). The AI economy centers elsewhere, on the question about individuals and their own data: Is gathered personal information processed to invigorate self-determination and expand opportunities, or does it suffocate and narrow possible human experiences?

For the question to be addressed by ethical finance, a customized model is needed for evaluating companies that operate with AI at their core. The model should identify investments that vitalize autonomy, while also supporting social wellbeing, and fortifying technical trustworthiness. The model should be intellectually robust, manageable for analysts, useful for portfolio managers, and credible for auditors. If those challenges are met, the contribution will be increased human freedom incarnated as investors empowered to act in accord with their own values.